\documentclass{ws-procs961x669}            
\begin{document}
\title{Cosmological and astrophysical results exploiting magnification bias with high-z sub-millimetre galaxies}

\author{L. Bonavera$^{1,2,a}$, M.M. Cueli$^{1,2,b}$, J Gonzalez-Nuevo$^{1,2,c}$}

\address{$^1$Physics Department, University of Oviedo,\\
Oviedo, 33007, Spain\\
$^2$Instituto Universitario de Ciencias y Tecnologías Espaciales de Asturias (ICTEA), \\
C. Independencia 13, 33004 Oviedo, Spain\\
$^a$bonaveralaura@uniovi.es - $^b$mcueli@uniovi.es - $^c$gnuevo@uniovi.es}



\begin{abstract}
The high-z submillimeter galaxies (SMGs) can be used as background sample for gravitational lensing studies thanks to their magnification bias, which can manifest itself through a non-negligible measurement of the cross-correlation function between a background and a foreground source sample with non-overlapping redshift distributions. In particular, the choice of SMGs as background sample enhances the cross-correlation signal so as to provide an alternative and independent observable for cosmological studies regarding the probing of mass distribution. 

In particular the magnification bias can be exploited in order to constrain the free astrophysical parameters of a Halo Occupation Distribution model and some of the main cosmological parameters. Urged by the improvements obtained when adopting a pseudo-tomographic analysis, It has been adopted a tomographic set-up to explore not only a $\Lambda$CDM scenario, but also the possible time evolution of the dark energy density in the $\omega_0$CDM and  $\omega_0\omega_a$CDM frameworks.
\end{abstract}


\bodymatter

\section{Magnification bias}\label{sec_magbias}

Magnification bias is due to the gravitational lensing effect and consists in the apparent excess in the number of high redshift sources (the lensed sources) close to the position of low redshift galaxies (the lenses). In fact, the gravitational lensing deflects the light rays of high redshift sources causing the stretching of the apparent sky area in the region affected by the lensing. This also causes a boost in the flux density of the high redshift sources, making them more likely to be detected above a given instrument flux density limit (e.g. Ref.~\citenum{SCH92}). This is sketched in Fig.~\ref{fig_magbias}(a). The red objects in the backward plane (source plane) are the sources with flux density limit above the instrument detection limit, and thus detectable while the orange ones are those with a flux density below the instrument detection limit, and thus not detectable unless boosted by the magnification bias. The blue objects in the lens plane are the possible lenses that might cause the magnification bias effect. Finally in the observer plane, the blue objects of the lens plane will be observable together with the red objects of the source plane and some of the orange objects whose flux have been boosted. The peculiarity of such magnified orange objects is that they will be found close to some of the blue objects (their lenses).

\def\figsubcap#1{\par\noindent\centering\footnotesize(#1)}
\begin{figure}[h]%
\begin{center}
\parbox{2.1in}
{\includegraphics[width=2in]{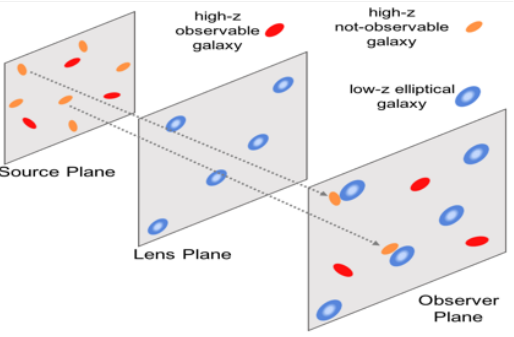}\figsubcap{a}}
\hspace*{4pt}
\parbox{2.1in}{\includegraphics[width=2in]{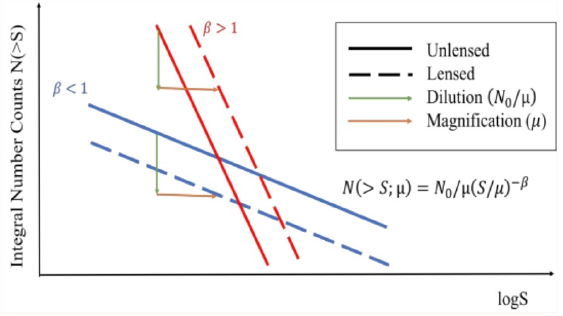}\figsubcap{b}}
\caption{Scheme of magnification bias (a) and how it affect the integral number counts of the lensed sources (b). See text for details.}
\label{fig_magbias}
\end{center}
\end{figure}

On this respect, magnification bias main advantage relies on the steepness of background sources counts and on the fact that it does not require either ellipticity measurements or knowledge on the galaxies’ orientation (as shear does). 
In particular, given that $n_0$ are the unlensed integrated background source number counts (number of background sources per solid angle and redshift with observed flux density larger than $S$ in the absence of gravitational lensing):
\begin{equation}
    n_0(>S,z)\equiv \int_S^{\infty} \frac{dN}{dSdz\,d\Omega}dS
\end{equation}
the background source number counts at an angular position $\vec{\theta}$ within an image are modified by dilution and magnification following Ref.~\citenum{BAR01}:
\begin{equation}
n(>S,z;\vec{\theta}\,)=\frac{1}{\mu(\vec{\theta}\,)}n_0\bigg(>\frac{S}{\mu(\vec{\theta}\,)},z\bigg)
\end{equation}
where $\mu(\vec{\theta})$ is the magnification field at angular position $\vec{\theta}$.
Assuming a power-law behaviour of the unlensed integrated number counts with $\beta$ the source number count steepness, i.e. $n_0(>S,z)=AS^{-\beta}$,  
\begin{equation}
\frac{n(>S,z;\vec{\theta})}{n_0(>S,z)}=\mu^{\beta-1}(\vec{\theta})\quad
\end{equation}
Depending on the value of $\beta$ the source counts might increase because of lensing: Fig.~\ref{fig_magbias}(b) sketched the increase or decrease of the source number counts for $\beta > 1$ (in red) and $\beta < 1$ (in blue), respectively.

\section{Cross-correlation}

Magnification bias is usually observed by means of angular cross-correlation function (CCF) measurements between two samples of sources: how many background sources fall ‘near’ the position of a lens, repeated for each object in the lens sample. Since the samples are at different redshifts, the detection of a non-zero signal would be exclusively due to the gravitational lensing induced by the lenses.

\subsection{The estimator}
Thus, magnification bias can be measured through the cross-correlation between the low and high redshift objects.

The CCF of two source populations ($D_f$ and $D_b$) is the fractional excess probability, relative to a random distribution ($R_f$ and $R_b$), of finding a $D_f$ source separated by an angle $\theta$ from a $D_b$ source within an infinitesimal solid angle $d\theta$. The CCF can be computed using a modified version of the Landy \& Szalay (Ref.~\citenum{LAN93}) estimator (see Ref.~\citenum{HER01}): 
\begin{equation}
    \tilde{w}_{fb}(\theta)=\frac{D_fD_b(\theta)-D_fR_b(\theta)-D_bR_f(\theta)+R_fR_b(\theta)}{R_fR_b(\theta)}
\end{equation}

In particular, $D_fD_b$, $D_fR_b$, $D_bR_f$ and $R_fR_b$ are the normalized foreground-background, foreground-Poisson, background-Poisson and Poisson-Poisson pair counts for a given angular separation $\theta$. 
Given two source samples with non-overlapping redshift distributions, the excess signal when computing their cross-correlation with respect to the random case is due to the magnification bias.  
Being such signal related to lensing and thus to cosmological distances and the galaxy halo characteristics, it can be used to constrain cosmological and astrophysical parameters.  

\subsection{The halo model}

To achieve this, the adopted theoretical description of the cross-correlation is the one by Cooray \& Sheth (Ref.~\citenum{COO02}): 
\begin{equation}
    w_{fb}(\theta)=2(\beta -1)\!\!\!\int^{\infty}_0\!\!\!\!\!\! \frac{dz}{\chi^2(z)}n_f(z)W^{\text{lens}}(z) \int_{0}^{\infty}\!\!\!\!l\frac{dl}{2\pi}{P_{\text{gal-dm}}(l/\chi^2(z),z)}J_0(l\theta)
\end{equation}
\begin{equation}
    W^{\text{lens}}(z)=\frac{3}{2}\frac{H_0^2}{c^2}\bigg[\frac{{E(z)}}{1+z}\bigg]^2\int_z^{\infty} dz' \frac{\chi(z)\chi(z'-z)}{\chi(z')}n_b(z')
\end{equation}
where $l$ is the multipole, $H_0$ is the Hubble constant, $E(z)$ the quantity where the contribution to the energy density are included (see section \ref{DE} for details), $n_b(z)$ ($n_f(z)$) is the normalized background (foreground) redshift distribution, $\chi(z)$ is the comoving distance, $J_0$ is the zeroth-order Bessel function of the first kind, $W^{\text{lens}}(z)$ is the lensing kernel and $\beta$ is the source number count steepness, commonly fixed to 3 for submillimeter galaxies (Ref.~\citenum{BON20}and reference therein). Therefore, the cross-correlation signal can be interpreted under the halo model parametrization (Ref.~\citenum{COO02}) considering that both galaxy samples trace the same dark matter distribution around redshift $z\sim0.4$. This dark matter distribution is traced directly by the foreground galaxies while, in the case of the background sample, it is traced thanks to the weak lensing effect. 
According to this model, the galaxy distribution power spectrum is parametrized as the sum of a 2-halo term, related to the correlations between one halo traced by the foreground galaxies and another one traced by the background sources and that therefore dominates at large scales, and a 1-halo term, that describes the correlation between sub-halos (traced by both samples) inside the same halo and therefore more important at small scales. Moreover, the halo model also suggests a simple parametrization of the cross-correlation between the galaxy and dark matter distributions (see Refs.~\citenum{COO02, SEL00,GUZ01}). 
According to the halo model (Cooray \& Sheth, Refs.~\citenum{COO02}):
\begin{equation}
P_{\text{g-dm}}(k,z)=P^{1h}_{\text{g-dm}}(k,z)+P^{2h}_{\text{g-dm}}(k,z)
\end{equation}

\begin{equation}
\begin{split}
    P_{\text{g-dm}}^{\text{1h}}(k,z)&=\int_0^{\infty} dM\,M\frac{n(M,z)}{\bar{\rho}(z)}\frac{\langle N_{g}\rangle_M}{\bar{n}_g(z)}|u_{\text{dm}}(k,z|M)||u_{\text{g}}(k,z|M)|^{p-1}\\
    P_{\text{g-dm}}^{\text{2h}}(k,z)&=P^{\text{lin}}(k,z)\Big[\int_0^{\infty}dM\,M\frac{n(M,z)}{\bar{\rho}(z)}b_1(M,z)u_{\text{dm}}(k,z|M)\Big]\,\cdot\nonumber\\
    &\quad\quad\quad\cdot\Big[\int_0^{\infty}dM\,n(M,z)b_1(M,z)\frac{\langle N_g \rangle_M}{\bar{n}_g(z)}u_g(k,z|M)\Big]
\end{split}
\end{equation}
In these equations $\overline{\rho}$ is the background density, $\langle N_{g}\rangle_M$ the mean number of galaxies within a halo of mass M, $\bar{n}_g(z)$ the mean number density of galaxies at redshift z, $p$ is set to 1 for central galaxies and to 2 for satellites (Ref.~\citenum{COO02}), $k$ is the wave number, $n(M,z)$ is the Sheth \& Tormen halo mass function by Ref.~\citenum{SHE99}, $P^{lin}(k,z)$ is the linear matter power spectrum, $b_1(M,z)$ is the linear large-scale bias (via the peak background split), and $u_{g}(k,z|M)$ is the normalized Fourier transform of the galaxy density distribution within a halo, which is assumed to equal the dark matter density profile, i.e. $u_{g}(k,z|M) = u_{dm}(k,z|M)$. Halos are defined as overdense regions whose mean density is 200 times the mean background density of the universe according to the spherical collapse model, and the halo density profiles $\rho(r)$ adopted is the one by Ref.~\citenum{NAV96}, hereafter NFW, with the concentration parameter in Ref.~\citenum{BUL01}. 

The halo occupation distribution (HOD) assumed is the one by Zheng et al. in Ref.~\citenum{ZHE05}:
\begin{equation}
    \label{eq:ncen01}
    N_\text{cen}(M_h) =
    \begin{cases}
    0 \quad \text{if}\ M_h < M_\text{min}\\
    1 \quad \text{otherwise}
    \end{cases}
\end{equation}
\begin{equation}
    \label{eq:nsat01}
    N_\text{sat}(M_h) = N_\text{cen}(M_h) \cdot \biggl(\dfrac{M_h}{M_1}\biggr)^{\alpha}
\end{equation}

In this model, the mean number of galaxies is represented by $N_{gal}$ where, the distinction between central and satellite galaxies is made: $N_{gal} = N_{cen} + N_{sat} = 1 + N_{sat}$. All halos above a minimum mass $M_{min}$ host a galaxy at their centre, while any remaining galaxies are classified as satellites and are distributed in proportion to the halo mass profile (see Ref.~\citenum{ZHE05}). Halos host satellites when their mass exceeds the $M_1$ mass, and the number of satellites is a power-law function of halo mass $N_\text{sat}(M_h)$. These parameters define the adopted HOD.

\subsection{The barotropic index of the dark energy}
\label{DE}
In the theoretical model, a possible redshift evolution of the dark energy (DE) can be introduced by allowing different values for the dark energy equation of state parameter, $\omega$.
For a flat cosmology, a redshift-dependent function $f(z)$ in the quantity
\begin{equation}
    E(z)\equiv \sqrt{\Omega_r(1+z)^4+\Omega_M(1+z)^3+\Omega_k(1+z)^2+{\Omega_{\text{DE}}f(z)}}
\end{equation}
takes into account the possible time evolution of the dark energy density, where $\Omega_{x}$ is the present-day density parameters for radiation $r$, matter ($M$), curvature ($k$) and dark energy $DE$. In particular, by assuming that the barotropic index of dark energy
\begin{equation}
    \label{eq:w}
    \omega(z)=\omega_0+\omega_a\frac{z}{1+z}
\end{equation}
the dark energy density has a redshift dependence given by $\rho(z)=\rho_0f(z)$, where
\begin{equation}
    f(z)=(1+z)^{3(1+\omega_0+\omega_a)}e^{-3\omega_a\frac{z}{1+z}}
\end{equation}
being $\rho_0$ the dark energy density at $z=0$. As for the dark energy density parameter, it can be written as
\begin{equation}
    \Omega_{\text{DE}}(z)=\frac{\Omega_{\text{DE}}f(z)}{E(z)^2}
\end{equation}
In this framework, the $\Lambda$CDM is recovered when $\omega_0=-1$ and $\omega_a=0$.

\section{Background and foreground samples}\label{sec_fb}

\def\figsubcap#1{\par\noindent\centering\footnotesize(#1)}
\begin{figure}[b]%
\begin{center}
\includegraphics[width=3in]{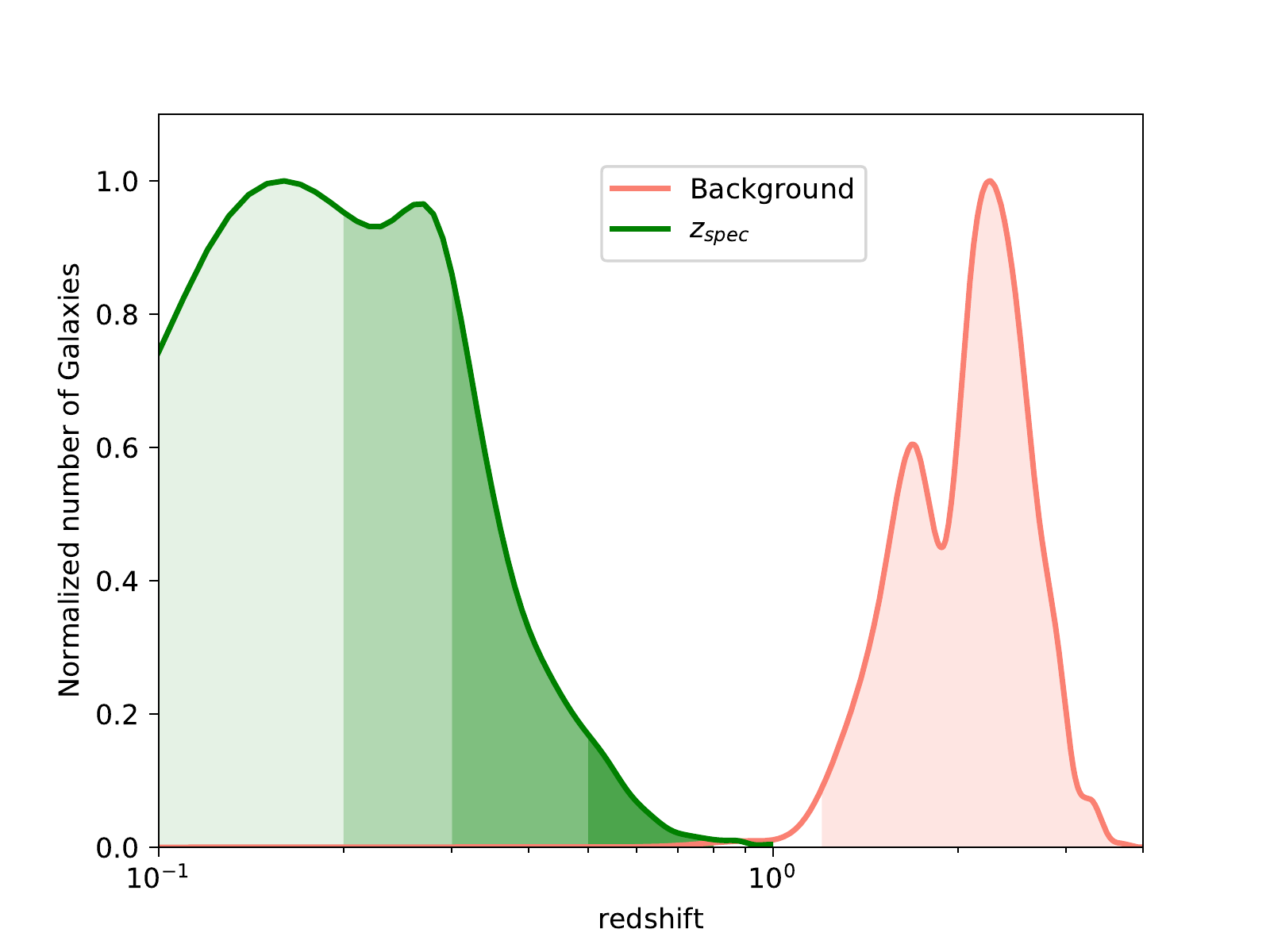}
\caption{Redshift distribution of background (orange) and foreground objects (green). The different shades of green indicate the bins of redshift used in Ref.~\citenum{BON21}. Credit: Bonavera et al., A\&A, in press, DOI 10.1051/0004-6361/202141521, 2021, reproduced with permission $\copyright$ ESO.}
\label{fig_zdistro}
\end{center}
\end{figure}

In order to achieve the cross-correlation measurement, a clear separation of the background and foreground sources is needed. For example, in the recent works by Bonavera et al. (Refs.~\citenum{BON20, BON21}), the redshift distribution of the foreground and background galaxies sample is the one in Fig.~\ref{fig_zdistro} in green and orange respectively. The different shades of green indicate the bins of redshift used in Ref.~\citenum{BON21}.

As for the foreground sample, different kind of sources at relatively low redshift can be used. 
For example, in their recent work (Ref.~\citenum{BON19}) Bonavera et al. adopt the same sample of quasi-stellar object (QSOs) with $0.2 < z < 1.0$ as in Ref.~\citenum{BIA19}. They are selected from the publicly available Sloan Digital Sky Survey (SDSS-II and SDSS-III) Baryon Oscillation Spectroscopic Survey (BOSS) catalogues. 
In Refs.~\citenum{GON17, BON20, GON21, CUE21, BON21} the galaxies in the Galaxy and Mass Assembly (GAMA, Ref.~\citenum{DRI11}) survey with $0.2 < z < 0.8$ have been used and in Ref.~\citenum{BON21} those with $0.1 < z < 0.8$ (see Fig.~\ref{fig_zdistro}).

Currently the best way to measure the CCF is to choose as background samples the high-$z$ sub-millimetre galaxies (SMGs). In fact they have steep  number counts $\beta \sim 3$ (Refs.~\citenum{BLA96, NEG07}), shown by the Herschel Space Observatory (Herschel; Ref.~\citenum{PIL10}) and the South Pole Telescope (SPT; Ref.~\citenum{CAR11}) observations. Such property enhances their magnification, as explained in Sect.~\ref{sec_magbias}.
Moreover, the SMGs are faint in the optical avoiding the possibility of being confused with the foreground lens sample, which is on its turn invisible at sub-millimetre (sub-mm) wavelength (Refs.~\citenum{AUG09,NEG10}). 
Finally, the redshifts of the SMGs are usually greater than $z > 1-1.5$, which guarantees no overlap with the foreground sample.
This makes the SMGs the perfect background sample for magnification bias studies by means of CCF measurements and thus allowing the tracing of the mass density (baryonic and dark matter) and its evolution with time. 

Moreover, it has been confirmed when, due to their magnification bias, Dunne et al. in Ref.~\citenum{DUN20} made a serendipitous direct observation of high-redshift SMGs. They were performing a study of gas tracers with Atacama Large Millimeter Array (ALMA) observation of galaxies by targeting a statistically complete sample of twelve galaxies selected at 250$\mu$m with z=0.35 and magnified SMGs appears around the position of half of them.  

\section{Recent results with SMGs magnification bias}

The first attempt to measure the CCF was carried out in Ref.~\citenum{WAN11} with Herschel/SPIRE galaxies and low-$z$ galaxies as background and foreground samples, respectively. Their results were a strong confirmation of this lensing-induced effect.

To study this possible bias, Ref.~\citenum{GON14} relied on much better statistics and carried out a more detailed analysis by measuring the CCF with Herschel Astrophysical Terahertz Large Area Survey (H-ATLAS) high-$z$ sources at $z > 1.5$ and two optical samples selected from SDSS (Sloan Digital Sky Survey, Ref.~\citenum{WAN11}) and GAMA (Ref.~\citenum{DRI11}) surveys, with redshifts $0.2 < z < 0.6$. The resulting CCF was measured with high significance, $ > 10 \sigma$ and using realistic simulations, they concluded that the signal was entirely explained with the magnification bias produced by the weak lensing effect caused by galaxy groups/clusters whose halo masses are in the range of $10^{13.2}-10^{14.5} M_{\odot}$ and that are signposted by the brightest galaxies in the optical samples. 
Later on, Gonazelz-Nuevo et al. in Ref.~\citenum{GON17} and Bonavera et al. in Ref.~\citenum{BON21} showed furthermore that the SMGs properties make them the perfect background sample for constraining the free parameters of a halo occupation distribution (HOD) model. 
In particular, the results by Ref.~\citenum{GON17} suggest that the lenses are massive galaxies or clusters, with a minimum mass of $M_{min}=10^{13} M_{\odot}$. They also carried out tomographic studies on the HOD parameters: they divided the foreground sample in four bins of redshift, $0.1 < z < 0.2$, $0.2 < z < 0.3$, $0.3 < z < 0.5$ and $0.5 < z < 0.8$. They main findings were that while $M_1$ is almost redshift independent, $M_{min}$ evolves: it increases with redshift, as predicted by theoretical estimations. 
Bonavera et al. in Ref.~\citenum{BON19} study with the magnification bias the mass properties of a sample of QSOs whose position signpost the lenses at $0.2<z<1.0$ obtaining $M_{min}= 10^{13.6^{+0.9}_{-0.4}} M_\odot$. This again suggests that the lensing is actually produced by halos of the size of clusters placed close to the QSOs positions.

Furthermore, magnification bias trough CCF measurements can be used to constrain the halo mass function. In particular, Cueli et al. in Ref.~\citenum{CUE21} successfully test such possibility according to two common parametrisations: the Sheth \& Tormen and Tinker fits. They find general agreement with traditional values for the involved parameters, with a slight difference in the Sheth \& Tormen fit for intermediate and high masses, where the results suggest a hint at a somewhat higher number of halos. 

\section{Cosmological studies with magnification bias}

Moreover, some of the main cosmological parameters can be also estimated using the magnification bias, as in Ref.~\citenum{BON20}. They use the CCF measured between a foreground sample of GAMA galaxies with spectroscopic redshifts in the range of $ 0.2 < z < 0.8$ and a background sample of H-ATLAS galaxies with photometric redshifts $z > 1.2$ (as described in section \ref{sec_fb}) to constrain the astrophysical parameters ($M_{min}$, $M_1$, and $\alpha$) and the $\Omega_M$, $\sigma_8$, and $H_0$ cosmological ones (the matter density parameter, the present root-mean-square matter fluctuation
averaged over a sphere of radius $8h^{–1}$Mpc where $h$ is the dimensionless Hubble constant, and the Hubble density parameters). These parameters are estimated through a Markov chain Monte Carlo analysis, using the Python package {\texttt{emcee}} (Ref.~\citenum{EMCEE}). They study various cases and in particular they perform a run by setting flat priors to those in table \ref{tab_bon20priors}.

\begin{table}
\tbl{Flat priors for the $\Lambda$CDM run in Ref.~\citenum{BON20}.}
{\begin{tabular}{@{}cccc@{}}
\toprule
log $M_{min}$ & $\mathcal{U}$[ 11.6, 13.6] & $\Omega_M$ & $\mathcal{U}$[  0.1,  0.8]\\
log $M_1$     &  $\mathcal{U}$[ 13.0, 14.5] & $\sigma_8$ & $\mathcal{U}$[  0.6,  1.2]\\
$\alpha$      &  $\mathcal{U}$[  0.5, 1.37] & $h$ & $\mathcal{U}$[  0.5,  1.0] \\\botrule
\end{tabular}}
\begin{tabnote} The HOD parameters names and flat priors are listed in the first two columns. The cosmological ones in the third and fourth columns.\\
\end{tabnote}\label{tab_bon20priors}
\end{table}

They obtain a lower limit at 95\% confidence level (CL) on $\Omega_M>0.24$, a slight trend towards $H_0>70$ km s$^{-1}$ Mpc$^{-1}$ values and an upper limit at 95\% CL of $\sigma_8<1$. Such results are summarised in the $\Omega_M-\sigma_8$ plane in Fig. \ref{fig_2Dplot} (a) with the dot-dashed blue line. 

Given the large number of studies that can be performed with magnification bias through CCF data, a systematic analysis of possible bias that might affect the estimation have been carried out. In particular, in Ref.~\citenum{GON21} Gonzalez-Nuevo et al. take into account different biases at cosmological scales in the source samples to carefully measure unbiased CCF. More precisely, their background sample consists of H-ATLAS galaxies with $z >1.2$ whereas they use  two independent foreground samples with $0.2< z <0.8$: GAMA galaxies with spectroscopic redshifts and SDSS galaxies with photometric redshifts. These independent samples allowed them to perform a pseudo-tomographic study that yields to constrain $\Omega_M= 0.50^{+0.14}_{-0.20}$ and $\sigma_8= 0.75^{+0.07}_{-0.10}$. Such analysis also suggests that a tomographic approach might improve the results.  

Driven by the conclusions mentioned above, Bonavera et al. in Ref.~\citenum{BON21} adopt the unbiased sample by Gonzalez-Nuevo et al. in Ref.~\citenum{GON21} dividing the foreground sample into four redshift bins ($0.1-0.2$, $0.2-0.3$, $0.3-0.5$ and $0.5-0.8$). The background objects are a sample of H-ATLAS galaxies with photometric redshifts $z > 1.2$. The redshift distribution are shown in Fig. \ref{fig_zdistro} and the redshift bins of the foreground sample are highlighted with different shades of green. They use magnification bias in tomography to jointly constrain the astrophysical HOD parameters ($M_{min}$, $M_1$ and $\alpha$) in each one of the selected redshift bins together with the $\Omega_M$, $\sigma_8$, and $H_0$ cosmological ones. In particular, the analysis is carried out both in the $\Lambda$CDM scenario and with the introduction of the dark energy density related $\omega_0$ and $\omega_a$ parameters in the $\omega_0$CDM and $\omega_0\omega_a$CDM frameworks.  
The parameters priors for the three described runs are listed in Table \ref{tab_bon21priors}.

\begin{table}
\tbl{Flat priors for the $\Lambda$CDM, $\omega_0$CDM and $\omega_0\omega_a$CDM models in Ref.~\citenum{BON21}.}
{\begin{tabular}{@{}cccc@{}}
\toprule
$\log{M_{min}} \mathcal{U}[a_i,b_i]$ & $\log{M_{1}}\mathcal{U}[c_i,d_i]$ & $\Omega_M$ & $\mathcal{U}$[0.1, 0.8] \\
    $[a_1,b_1]$ = $\mathcal{U}[10.0,13.0]$ & $[c_1,d_1]$ = $\mathcal{U}[12.0,15.5]$ & $\sigma_8$ & $\mathcal{U}$[0.6, 1.2]\\
    $[a_2,b_2]$ = $\mathcal{U}[11.0,13.0]$ & $[c_2,d_2]$ = $\mathcal{U}[12.0,15.5]$ & $h$ & $\mathcal{U}$[0.5, 1.0]\\
    $[a_3,b_3]$ = $\mathcal{U}[11.5,13.5]$ & $[c_3,d_3]$ = $\mathcal{U}[12.5,15.5]$ & $\omega_a$ & $\mathcal{U}$[ -3.0, 3.0]\\
    $[a_4,b_4]$ = $\mathcal{U}[13.0,15.5]$ & $[c_4,d_4]$ = $\mathcal{U}[13.0,15.5]$ & $\omega_0$ & $\mathcal{U}$[ -2.0,  0.0] \\\botrule
\end{tabular}}
\begin{tabnote} For the four different bins of redshift, the $M_{min}$ flat priors are listed in the first column while the $M_1$ ones in the second column. The $\alpha$ prior is $\alpha = \mathcal{U}$[ 0.5, 1.37] in all bins. The cosmological parameters names and flat priors are in the third and fourth columns.\\
\end{tabnote}\label{tab_bon21priors}
\end{table}

\def\figsubcap#1{\par\noindent\centering\footnotesize(#1)}
\begin{figure}[h]%
\begin{center}
\parbox{2.1in}
{\includegraphics[width=2in]{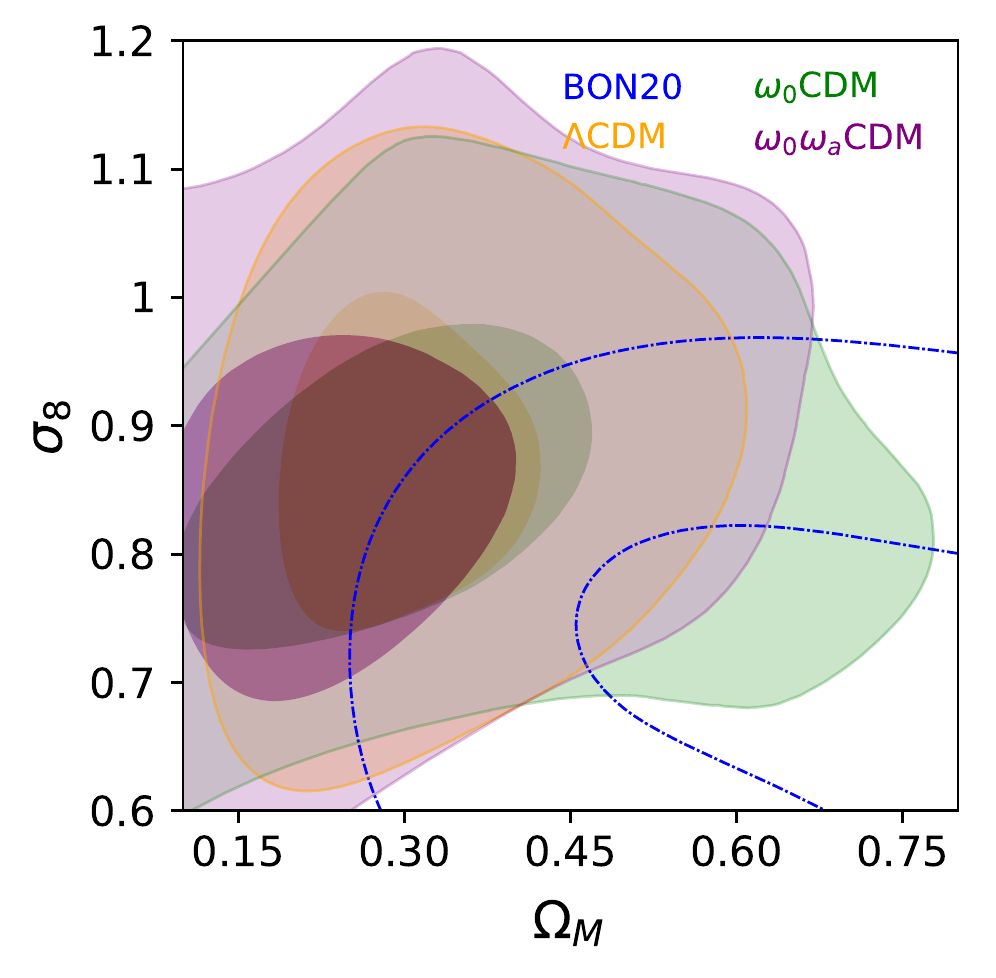}\figsubcap{a}}
\hspace*{4pt}
\parbox{2.1in}{\includegraphics[width=2in]{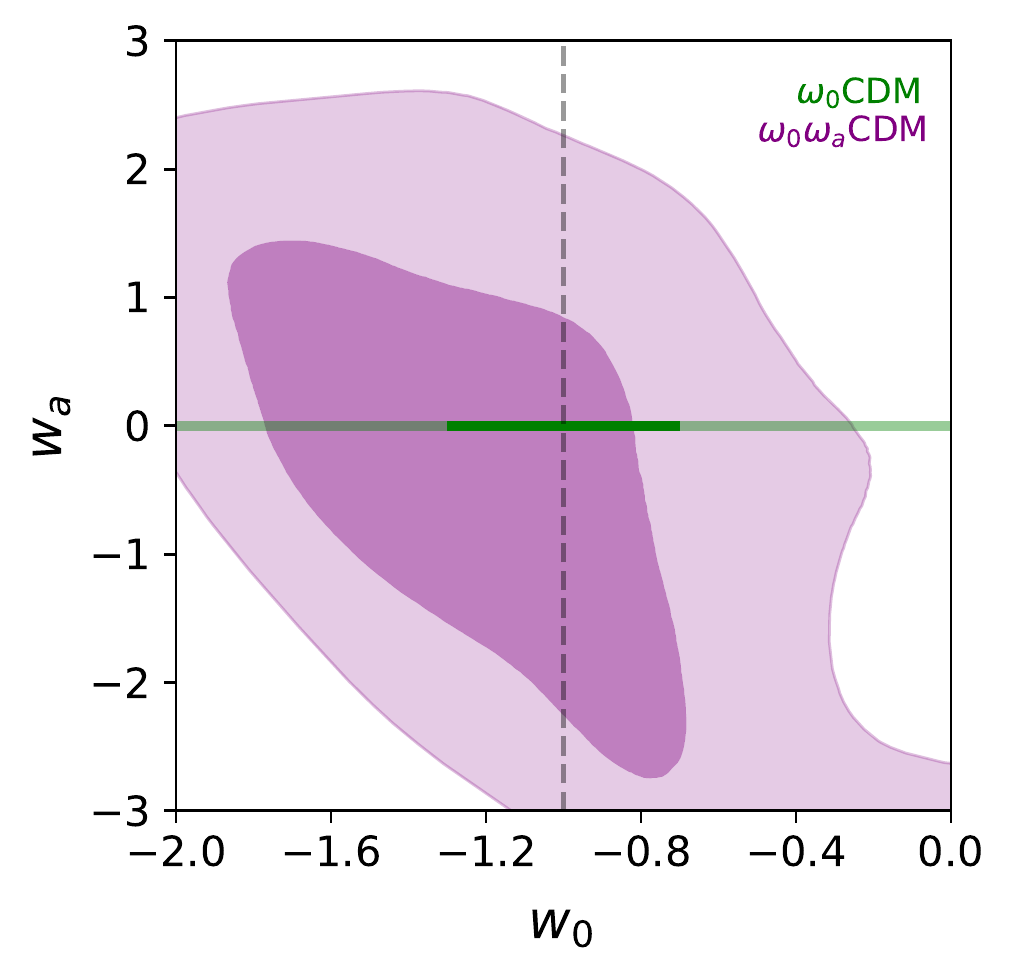}\figsubcap{b}}
\caption{Contour plot of the 2D posterior distributions with contours set to 0.393 and 0.865. (a) $\Omega_M$ and $\sigma_8$ for the non-tomographic case in Ref.~\citenum{BON20} (dot-dashed blue line) and the tomographic runs in Ref.~\citenum{BON21} within $\Lambda$CDM (in yellow), $\omega_0$CDM (in green) and $\omega_0\omega_a$CDM (in purple). (b) $\omega_0$ and $\omega_a$ for the $\omega_0\omega_a$CDM model. Credit: Bonavera et al., A\&A, in press, DOI 10.1051/0004-6361/202141521, 2021, reproduced with permission $\copyright$ ESO.}%
\label{fig_2Dplot}
\end{center}
\end{figure}

As for the HOD parameters, $M_{min}$ shows a trend towards higher values at higher redshift confirming the findings in Ref.~\citenum{GON17}. For the $\Lambda$CDM model, they obtain a mean(maximum) posterior value [68\% CL] of $\Omega_M=0.33(0.26)$ [0.17,0.41] and of $\sigma_8$= 0.87 [0.75,1], being $H_0$ not yet constrained. For the $\omega_0$CDM model, they find similar results on $\Omega_M$ and $\sigma_8$ and a mean(maximum) posterior value [68\% CL] of $\omega_0=-1(-0.97)$ $[-1.56, -0.47]$. 
For the $\omega_0\omega_a$CDM model, $\omega_0=-1.09(-0.92)$ $[-1.72,-0.66]$ and $\omega_a=-0.19(-0.20)$ $[-1.88,1.48]$. 
These results are summarised in Tables \ref{tab_resultsLCDM} and \ref{tab_results}, where the the mean, peak and $68\%CL$ of the posterior distributions for the estimated cosmological parameters are given. In particular, the first column gives the parameter name, the second and third columns in Table \ref{tab_resultsLCDM} are the results for the $\Lambda$CDM model in the non-tomographic and tomographic cases, and in Table \ref{tab_results} are those for the $\omega_0$CDM and $\omega_0\omega_a$CDM models in the tomographic case.

\begin{figure}[h]%
\begin{center}
\parbox{2.1in}
{\includegraphics[width=3in]{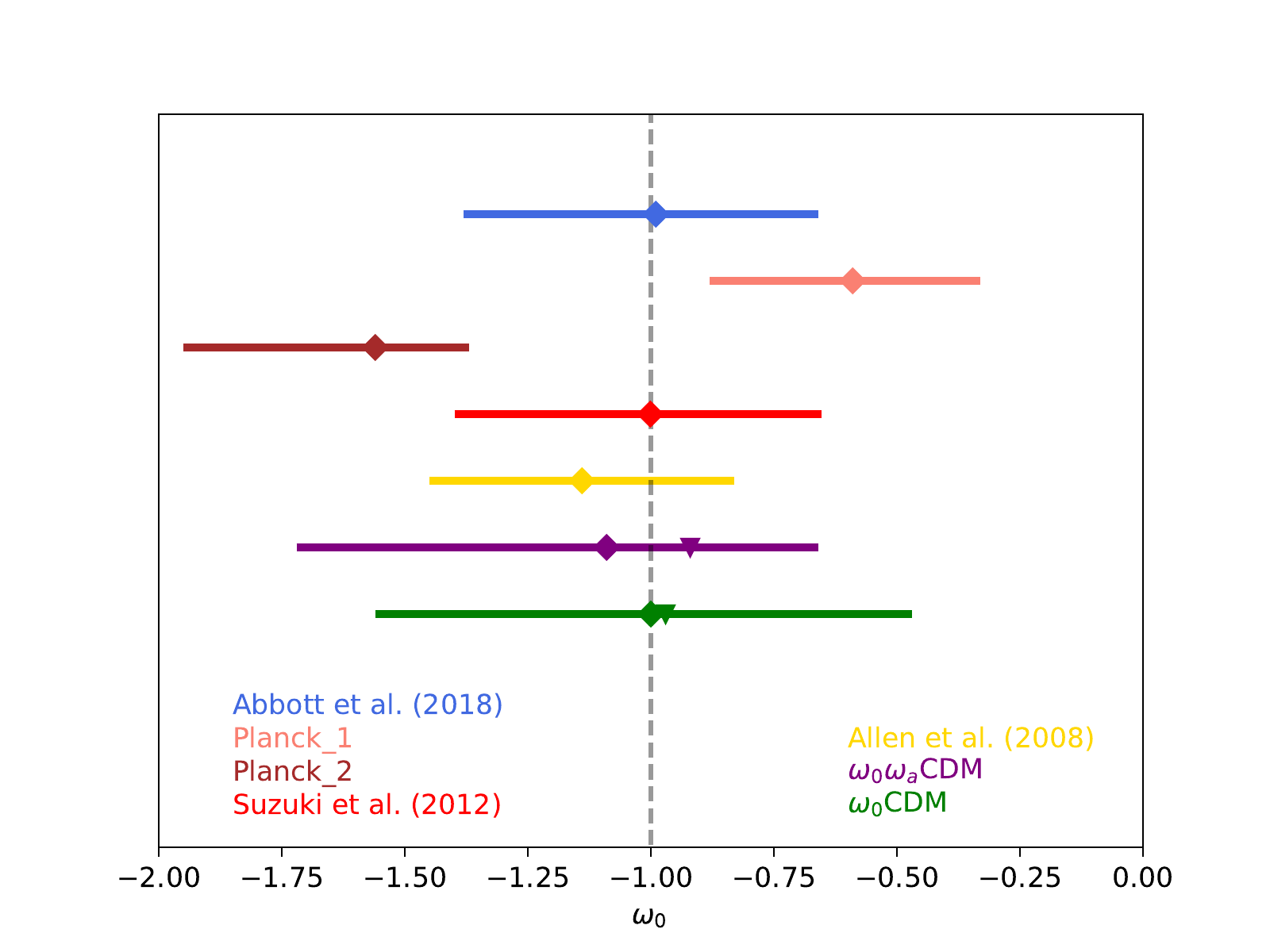}}
\caption{Results on $\omega_0$ within the $\omega_0$CDM (green) and $\omega_0\omega_a$CDM (purple) models compared with those by DES (blue), Planck1 (basewwaplikHMTTlowllowEBAO, salmon), Planck2 (basewplikHMTTlowllowE, brown), Supernovae (red), x-ray measurements (yellow). Credit: Bonavera et al., A\&A, in press, DOI 10.1051/0004-6361/202141521, 2021, reproduced with permission $\copyright$ ESO.}%
\label{fig_w0}
\end{center}
\end{figure}

The $\Omega_M-\sigma_8$ plane is shown in Fig.~\ref{fig_2Dplot}(a) for the $\Lambda$CDM (in yellow), $\omega_0$CDM (in green) and $\omega_0\omega_a$CDM (in purple) models. The results in the $\omega_0-\omega_a$ plane are shown in Fig.~\ref{fig_2Dplot}(b). The $\omega_0$ results are shown in Fig. \ref{fig_w0} for the $\omega_0$CDM (in green) and $\omega_0\omega_a$CDM (in purple) cases, where they are compared with other results from literature. Moreover, the tomographic analysis presented in Ref.~\citenum{BON21} confirms that magnification bias results do not show the degeneracy found with cosmic shear measurements and that, related to dark energy, do show a trend of higher $\omega_0$ values for lower $H_0$ values.  

\begin{table}
\tbl{Cosmological results for the $\Lambda$CDM from Bonavera et al. in Refs.~\citenum{BON20,BON21}.}
{\begin{tabular}{@{}ccccc@{}}
\toprule
 & \multicolumn{2}{c}{$\Lambda$CDM nontomo} & \multicolumn{2}{c}{$\Lambda$CDM} \\ 
           & $\mu(Peak)$ & $68\%CL$    & $\mu(Peak)$  & $68\%CL$           \\\colrule
$\Omega_M$ & 0.54(0.67)  & [0.46,0.80] & 0.33(0.26) &  [0.17,0.41]             \\   
$\sigma_8$ & 0.78(0.74)  & [0.63,0.85] & 0.87(0.87) &  [0.75,1.00]             \\   
$h$        & 0.76(-)     & [0.68,1.00] & 0.72(0.72) &  $<$0.79                 \\\botrule
\end{tabular}}
\begin{tabnote} Mean(Peak) and $68\%CL$ of the posterior distributions for the cosmological parameters (listed in the first column) estimated according to the $\Lambda$CDM model in the non-tomographic case (second and third columns) and in the tomographic one (fourth and fifth columns) models in the tomographic case. With the exception of the estimated cosmological parameters, the cosmology is fixed to the \textit{Planck} one.\\
\end{tabnote}\label{tab_resultsLCDM}
\end{table}

\begin{table}
\tbl{Cosmological results for the $\omega_0$CDM and $\omega_0\omega_a$CDM models from Bonavera et al. in Ref.~\citenum{BON21}.}
{\begin{tabular}{@{}ccccc@{}}
\toprule
  & \multicolumn{2}{c}{$\omega_0$CDM} & \multicolumn{2}{c}{$\omega_0\omega_a$CDM}\\ 
           & $\mu(Peak)$   & $68\%CL$      & $\mu(Peak)$   & $68\%CL$        \\\colrule
$\Omega_M$ &   0.38(0.26)  &  [0.13,0.47]    &  0.34(0.21)   &  [0.11,0.41]      \\   
$\sigma_8$ &   0.87(0.85)  &  [0.73,0.98]    &  0.88(0.84)   &  [0.72,1.01]      \\   
$h$        &   0.70(-)     &  $<$0.75        &  0.70(-)      &  $<$0.76          \\
$\omega_0$ & -1.00(-0.97)  &  [-1.56,-0.47]  & -1.09(-0.92)  & [-1.72,-0.66]     \\
$\omega_a$ &  -            &  -              & -0.19(-0.20)  & [-1.88,1.48]      \\\botrule
\end{tabular}}
\begin{tabnote} Mean(Peak) and $68\%CL$ of the posterior distributions for the cosmological parameters (listed in the first column) estimated according to the $\omega_0$CDM (second and third columns) and $\omega_0\omega_a$CDM (fourth and fifth columns) models in the tomographic case.  With the exception of the estimated cosmological parameters, the cosmology is fixed to the \textit{Planck} one.\\
\end{tabnote}\label{tab_results}
\end{table}

\def\figsubcap#1{\par\noindent\centering\footnotesize(#1)}

\section{Conclusion}

In conclusion, all these works confirm that the SMGs are a perfect background sample for magnification bias studies and serendipitous direct observation of magnified SMGs have been performed with ALMA (i.e. Ref.~\citenum{DUN20}). It has been demonstrated in Ref.~\citenum{GON14} and Ref.~\citenum{GON17} for astrophysical studies and in Ref.~\citenum{BON20} for cosmological ones. In particular magnification bias proves to be useful as an independent and additional cosmological probe for $\Lambda$CDM and beyond $\Lambda$CDM models, as done in Ref.~\citenum{BON21}. 

In these works, the performance of magnification bias through CCF measurements has been compared in non-tomographic and tomographic analysis. The conclusion is that tomography improves non-tomographic studies despite the worsening in the statistics of the measured CCFs. 

Moreover, the CCF can be computed adopting different foreground samples, e.g. galaxies in Ref.~\citenum{BON20} or QSOs in Ref.~\citenum{BON19} and the measurements are improved by taking into account observational biases correction (i.e. Ref.~\citenum{GON21}). In addition, variation of the ingredients of the models, as the HMF, might be tested with magnification bias (e.g. Ref.~\citenum{CUE21}).

Up to now, the main conclusions are the fact that astrophysical results are robust for all models ($\Lambda$CDM and beyond $\Lambda$CDM) and that $\Lambda$CDM compatible results have been found with tomographic and non-tomographic approaches, i.e. $\Omega_M \sim 0.33$, $\sigma_8 \sim 0.87$, $\omega_0 \sim -1$ and $\omega_a \sim -0.19$.

\section*{Acknowledgments}
LB, JGN and MMC acknowledge the PGC 2018 project PGC2018-101948-B-I00 (MICINN/FEDER). MMC acknowledges support from PAPI-20-PF-23 (Universidad de Oviedo). LB acknowledges Carlo Burigana forhis help in the clarity of the manuscript.

\bibliographystyle{ws-procs961x669}
\bibliography{magbias}

\end{document}